\documentclass{elsart4}

\usepackage{graphicx}

\usepackage{amssymb,amsbsy,amsmath}

\newcommand{\op}[1]{%
    \fontdimen12\textfont3=2pt\fontdimen12\scriptfont3=1.4pt%
    \!\null\mathop{\vphantom{#1}\smash{#1}}\limits_{\sim}\null\!}
\newcommand{\xref}[1]{\protect\ref{#1}}
\newcommand{\figref}[1]{Fig.~\protect\ref{#1}}
\newcommand{\fmref}[1]{(\protect\ref{#1})}
\def\bra#1{\langle \, {#1} \, | \,}
\def\ket#1{\, | \, {#1} \, \rangle}

\begin{document}
\begin{frontmatter}

  \title{Theoretical estimates for proton-NMR spin-lattice
    relaxation rates of heterometallic spin rings}


\author{Mohammed Allalen and J\"urgen Schnack\corauthref{cor1}}
\address{Universit\"at Osnabr\"uck, Fachbereich Physik,
D-49069 Osnabr\"uck, Germany}
\corauth[cor1]{Tel: ++49 541 969-2695; fax: -12695; Email: jschnack@uos.de}

\begin{abstract}
  Heterometallic molecular chromium wheels are fascinating new
  magnetic materials. We reexamine the available experimental
  susceptibility data on {MCr$_7$} wheels in terms of a simple
  isotropic Heisenberg Hamiltonian for M=Fe, Ni, Cu, and Zn and
  find in that {FeCr$_7$} needs to be described with an
  iron-chromium exchange that is different from all other cases.
  In a second step we model the behavior of the proton spin
  lattice relaxation rate as a function of applied magnetic
  field for low temperatures as it is measured in Nuclear
  Magnetic Resonance (NMR) experiments.  It appears that
  {CuCr$_7$} and {NiCr$_7$} show an unexpectedly reduced
  relaxation rate at certain level crossings.
\end{abstract}

\begin{keyword}
\PACS 75.10.-b\sep 75.10.Jm\sep 75.50.Ee
\KEY  Heisenberg model \sep Molecular magnets\sep Spin rings \sep  Antiferromagnetism
\end{keyword}
\end{frontmatter}

\section{Introduction}
\label{sec-1}

Among magnetic molecules spin rings constitute a rich subgroup
of highly symmetric species of various sizes which are
comprising a large variety of paramagnetic ions
\cite{TDP:JACS94,ACC:ICA00,WKS:IO01,VSG:CEJ02}.  The
investigation of these regular structures led to a deeper
understanding especially of antiferromagnetically coupled spin
systems. One of the findings is the discovery and confirmation
of rotational bands, see e.g.
\cite{ScL:PRB01,Wal:PRB02,Waldmann:EPL02}.

In accord with these investigations it was anticipated that spin
rings, which host an odd number of spins or spins of different
size, would show complementary quantum effects that would be
interesting on their own. Odd membered rings for instance would
violate the presuppositions for the theorems of Lieb, Schultz,
and Mattis \cite{LSM:AP61,LiM:JMP62,Wal:PRB02} and thus possess
non-trivial ground states as well as low-lying excited states
\cite{BHS:PRB03} with quantum numbers and degeneracies that
differ from those of their bipartite, i.e. even-membered
counterparts. The N\'eel-like local magnetization which in
even-membered rings results from a superposition of the singlet
ground state and the $(M=0)$-component of the first excited
triplet state \cite{HML:EPJB02,MeL:PB03} would in odd-membered
rings assume the form of a topological soliton \cite{ScS:JMMM05}
that equally well could be pictured as a M\"obius strip
\cite{CGS:JMMM05}.  Although it is rather difficult to
synthesize homometallic odd-membered rings, the prospects of
interesting features due to frustration nevertheless fuel future
efforts to synthesize odd rings.

In the case of spin ring systems comprising ions of different
chemical elements the breakthrough was already achieved with the
synthesis of heterometallic {MCr$_7$} wheels \cite{LME:ACIE03},
where one of the chromium ions of the original Cr$_8$ ring
\cite{VSG:CEJ02,CVG:PRB03,AGC:PRB03,AGC:JMMM04} is replaced by
another element M=Mn, Fe, Co, Ni, Cu, Zn, and Cd.  The
possibility of a systematic study has initiated first
investigations on these compounds as there are susceptibility
measurements \cite{LME:ACIE03} as well as neutron scattering on
{MnCr$_7$}, {ZnCr$_7$}, and {NiCr$_7$} wheels \cite{CGA:04}.

In this article we reexamine earlier susceptibility measurements
\cite{LME:ACIE03} by means of complete diagonalization
(Sec.~\xref{sec-2}) in the framework of an isotropic Heisenberg
model. Our results agree with first estimates given in
Ref.~\cite{LME:ACIE03} with the noticeable difference that we
find that the exchange parameters of the iron ion to its
neighboring chromium ions in the {FeCr$_7$} wheel is rather
different from the original chromium-chromium exchange whereas
it remains practically unchanged for the other paramagnetic
ions \cite{LME:ACIE03}.

In a second step (Sec.~\xref{sec-3}) we investigate the
principle structure of the proton spin-lattice relaxation rate
$T_1^{-1}$ as a function of the applied magnetic field strength
at low temperatures. This quantity can be probed by Nuclear
Magnetic Resonance (NMR). NMR has shown to be a powerful tool to
investigate the local spin dynamics in magnetic molecules
especially in the vicinity of level crossings
\cite{JJL:PRL99,LBJ:JMMM04,BLL:PRB04}. The relaxation rate
$T_1^{-1}$ is expected to increase drastically whenever two
levels approach each other due to possible resonant energy
exchange with the surrounding protons. In contrast to this
expectation it appears that {CuCr$_7$} and {NiCr$_7$} show an
unexpectedly reduced relaxation rate at certain level crossings
which should experimentally be observable.

The article closes with a summary and an outlook in
Sec.~\xref{sec-5}.

\section{Heisenberg Hamiltonian}
\label{sec-2}

The Hamilton operator of the isotropic Heisenberg model for
heterometallic {MCr$_7$} wheels is given by
\begin{eqnarray}
\label{E-2-1}
\op{H}
&=&
2\,J_1\,
\sum_{i=1}^6\;
\op{\vec{s}}(i) \cdot \op{\vec{s}}(i+1)
\\
&&
+
2\,J_2\,
\left(
\op{\vec{s}}(7) \cdot \op{\vec{s}}(8)
+
\op{\vec{s}}(8) \cdot \op{\vec{s}}(1)\right)
\nonumber
\ .
\end{eqnarray}
$J_1$ denotes the exchange parameter between nearest neighbor
chromium ions whereas $J_2$ denotes the exchange parameters
between the dopant and the two neighboring chromium ions.  We
chose $J>0$ for antiferromagnetic interaction in this article.

Neglecting anisotropy the Hamiltonian commutes with the square
$\op{\vec{S}}^2$ and the $z$-component $\op{S}_z$ of the total
spin. In addition point group symmetries can usually be
exploited.  In the following cases of heterometallic {MCr$_7$}
wheels only the mirror symmetry about the dopant is used. Then
for not too large subspaces all energy eigenvalues and
eigenvectors can be computed.

\section{Low-field susceptibility}
\label{sec-3}

\begin{figure}[!ht]
\begin{center}
\includegraphics[clip,width=60mm]{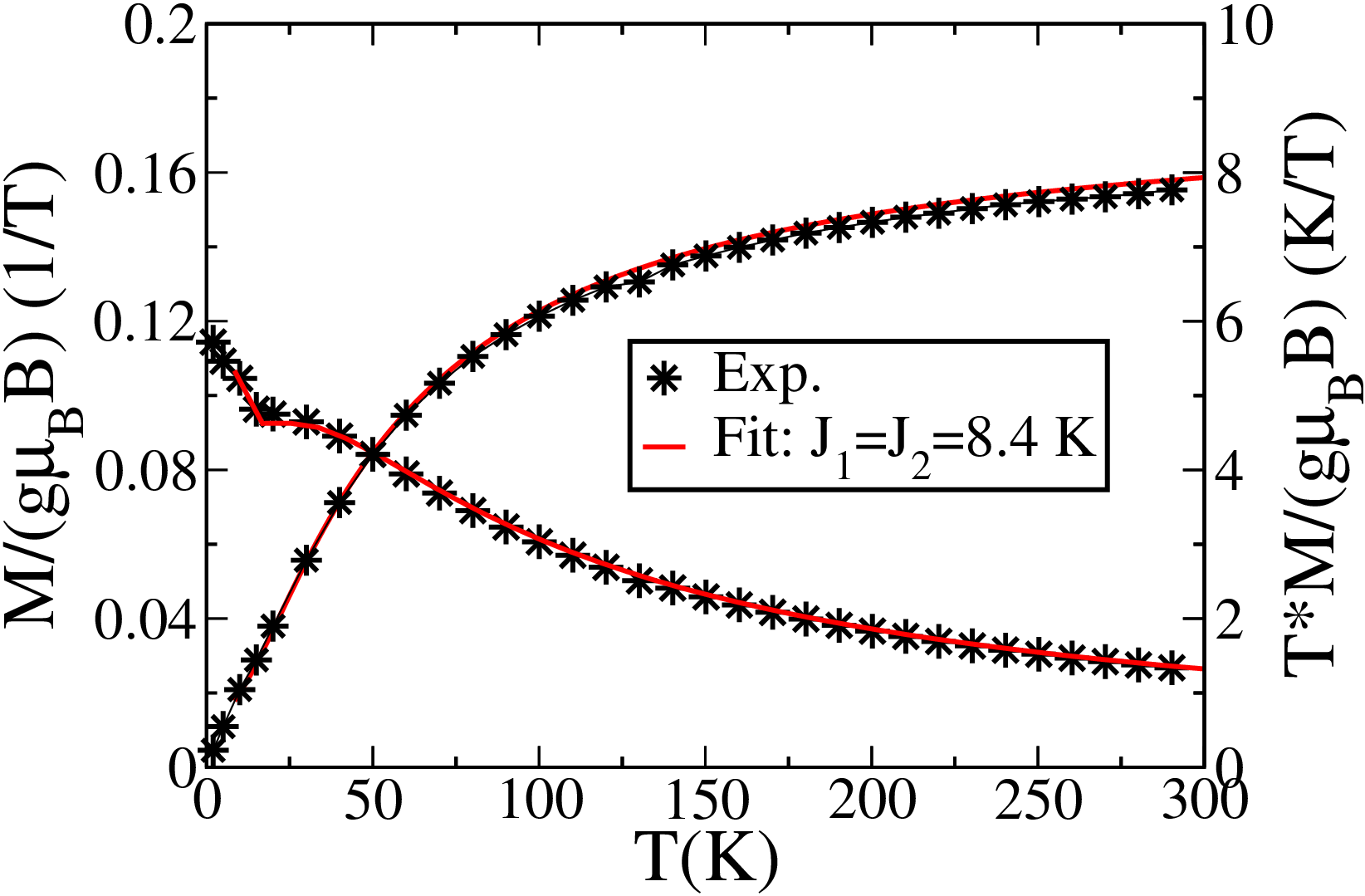}
\vspace*{1mm}
\caption[]{Variation of $\mathcal{M}/B$ and $T\mathcal{M}/B$ as
  a function of temperature $T$ for CuCr$_{7}$: The experimental
  data are given by black stars. The theoretical fit is depicted
  by a solid curve for $J_1=J_2=8.4$~K; $B=1$~T and $g=2$.}
\label{F-C}
\end{center}
\end{figure}

\begin{figure}[!ht]
\begin{center}
\includegraphics[clip,width=60mm]{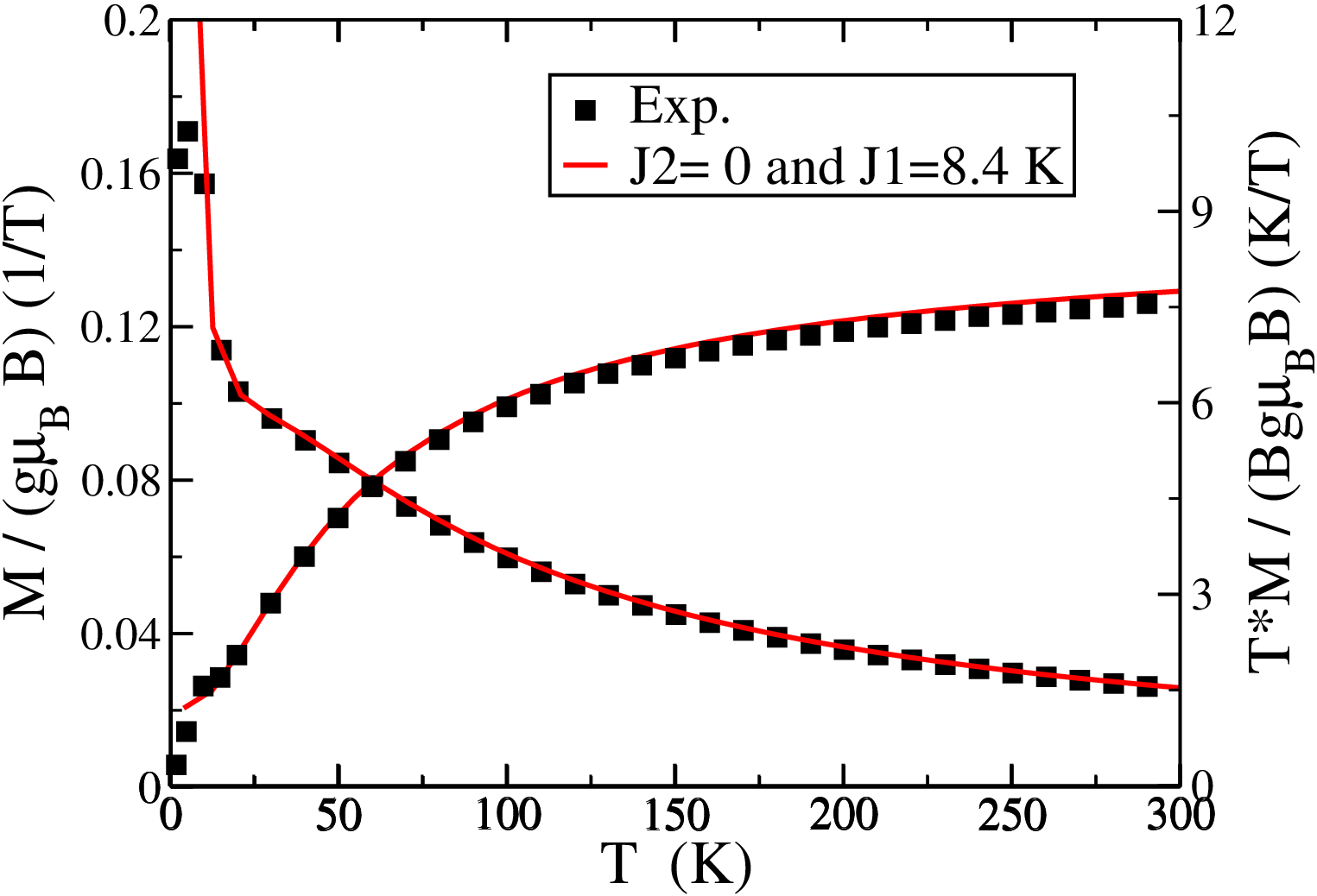}
\vspace*{1mm}
\caption[]{Variation of $\mathcal{M}/B$ and $T\mathcal{M}/B$ as
  a function of temperature $T$ for ZnCr$_{7}$: The experimental
  data are given by black squares. The theoretical fit is
  depicted by a solid curve for$J_1=8.4$~K and $J_2=0$~K;
  $B=1$~T and $g=2$.}
\label{F-D}
\end{center}
\end{figure}

\begin{figure}[!ht]
\begin{center}
\includegraphics[clip,width=60mm]{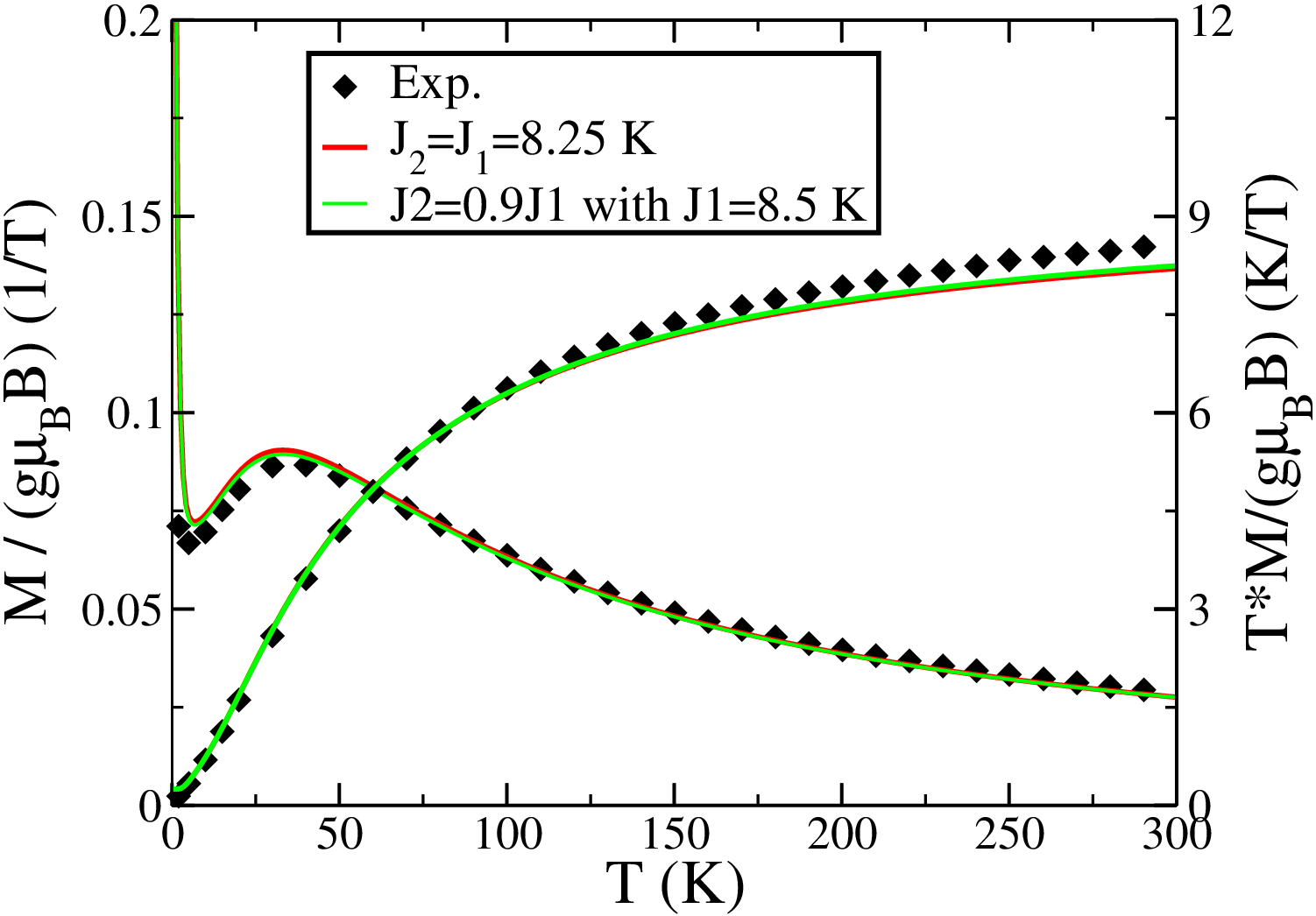}
\vspace*{1mm}
\caption[]{Variation of $\mathcal{M}/B$ and $T\mathcal{M}/B$ as
  a function of temperature $T$ for NiCr$_{7}$: The experimental
  data are presented as black diamonds. The theoretical fit for
  $J_1=J_2=8.25$~K is given by a solid curve and for $J_1=8.5$~K
  and $J_2=7.425$~K by a dashed curve.  $B=1$~T and $g=2.1$.}
\label{F-E}
\end{center}
\end{figure}

Reexamining the available experimental susceptibility data
\cite{LME:ACIE03} in terms of complete numerical diagonalization
of Hamiltonian \fmref{E-2-1} we find the same qualitative
behavior as in Ref.~\cite{LME:ACIE03}. Depending on the spin of
the dopant the resulting ground state spin $S$ assumes the
following values: $S=1/2$ for M=Fe, $S=1$ for M=Cu, $S=1/2$ for
M=Ni, and $S=3/2$ for M=Zn, compare also \cite{CGA:04}.

\begin{figure}[!ht]
\begin{center}
\includegraphics[clip,width=60mm]{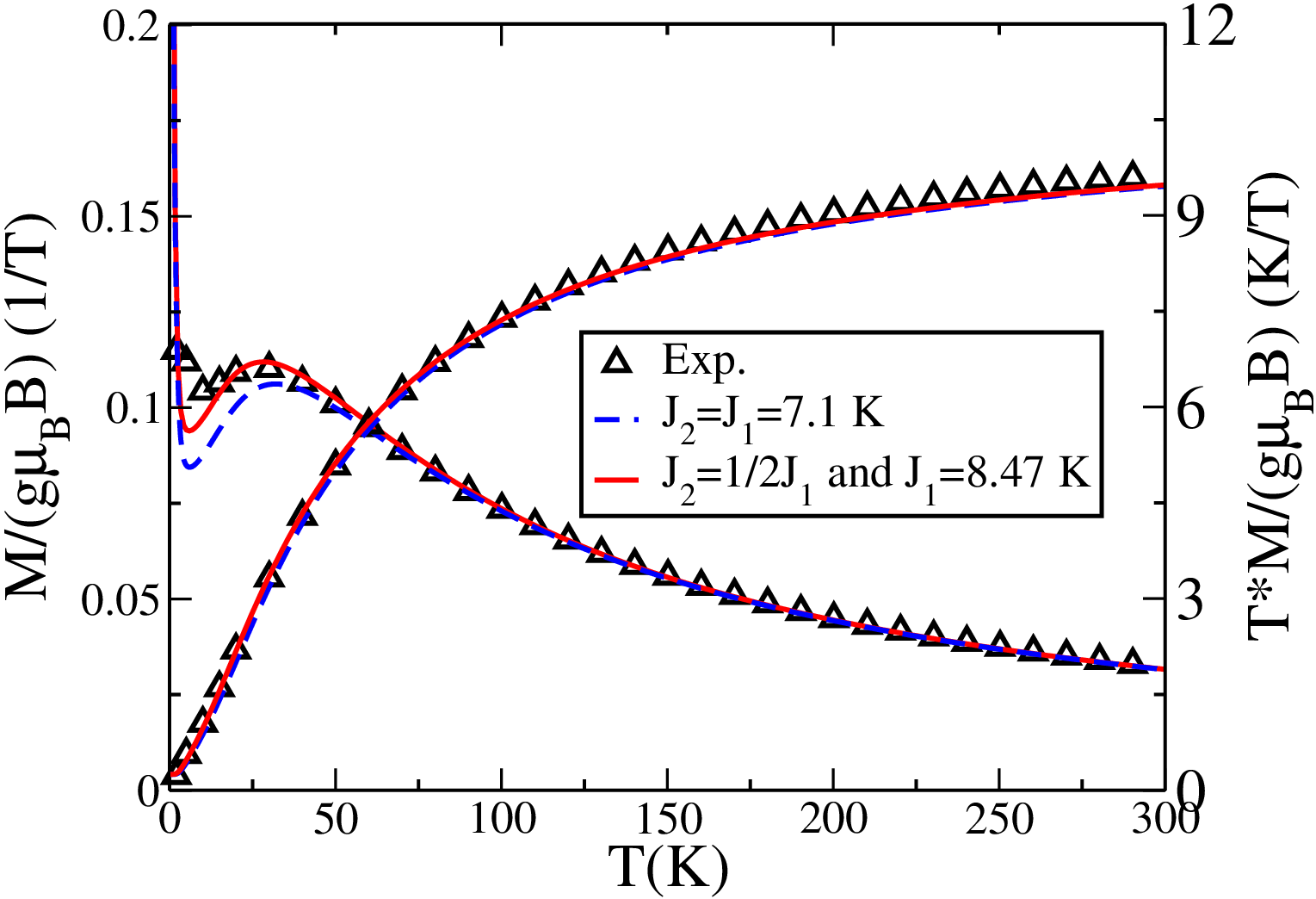}
\vspace*{1mm}
\caption[]{Variation of $\mathcal{M}/B$ and $T\mathcal{M}/B$ as
  a function of temperature $T$ for FeCr$_{7}$: The experimental
  data are depicted by black triangles. The theoretical fit for
  $J_1=J_2=7.1$~K is given by a dashed curve and for
  $J_1=8.47$~K and $J_2=J_1/2$ by a solid curve.  $B=1$~T and
  $g=2$.}
\label{F-B}
\end{center}
\end{figure}

The susceptibility $\mathcal{M}/B$ as well as $T\mathcal{M}/B$
of {CuCr$_7$}, {ZnCr$_7$}, {NiCr$_7$}, and {FeCr$_7$} are shown
in Figures~\xref{F-C}~-~\xref{F-B}. For the theoretical fits a
$g$-value of $g=2.1$ has been used for NiCr$_{7}$, in all other
cases $g=2$ has been assumed.

All susceptibility curves are compatible with antiferromagnetic
exchange. In the first example of {CuCr$_7$}, \figref{F-C}, a
common exchange interaction explains the experimental data. This
exchange is practically the same as in Cr$_8$ \cite{VSG:CEJ02},
thus unchanged in the heterometallic compound. The second
example of {ZnCr$_7$}, \figref{F-D}, constitutes a spin chain
since the Zn ion is diamagnetic. The original Cr-Cr interaction
is not altered whereas the coupling to the Zn ion is $J_2=0$.
The third example deals with {NiCr$_7$}, \figref{F-E}.  Here we
find that the experimental data can either be described by a
common but slightly reduced exchange interaction or by an almost
unchanged Cr-Cr interaction and a 10~\% smaller Cr-Ni exchange.
This has also been reported in Ref.~\cite{TGA:PRL05}, whereas
Ref.~\cite{CGA:04} favors a 15~\% bigger $J_2$.  The last
example of {FeCr$_7$}, \figref{F-B}, shows the biggest deviation
from the assumption of a common and almost unchanged exchange
parameter.  Although a single exchange constant provides a
reasonable fit to the experimental data \cite{LME:ACIE03}, a
better approximation -- especially at low temperatures -- is
given if one assumes that the Cr-Cr exchange is not much altered
whereas the Cr-Fe exchange is reduced to half the size of the
Cr-Cr exchange.

\section{Spin-lattice relaxation rates}
\label{sec-4}

Having determined the Heisenberg exchange parameters of several
heterometallic {MCr$_7$} wheels we investigate how the rather
different structures of low-lying levels of the various rings
expresses itself in proton spin-lattice relaxation rates as
would be measured by Nuclear Magnetic Resonance (NMR).

Following the general theory of nuclear relaxation
\cite{Abr:CO61} we determine the inverse relaxation time
$T_1^{-1}$ from spin-spin correlation functions as 
\begin{eqnarray}
\nonumber
\frac{1}{T_1}
&=&
\left(1+e^{-\frac{\hbar\omega_N}{k_BT}}\right) \frac{2\pi}{Z(T,B)}
\sum_{\mu,\nu} e^{-\beta E_\mu}
\bra{\psi_\mu}\op{F}^+ \ket{\psi_\nu}
\\
&&\times 
\bra{\psi_\nu}\op{F}^- \ket{\psi_\mu} 
\delta_\varepsilon(\omega_N - \frac{E_{\mu}-E_\nu }{\hbar})
\label{E-3-1}
\ .
\end{eqnarray}
Here $\omega_N$ denotes the nuclear Lamor frequency, $E_{\mu}$
and $E_{\nu}$ are energy eigenvalues of Hamiltonian
\fmref{E-2-1} augmented by a Zeeman term. The operators
$\op{F}^{\pm}$ are given by
\begin{eqnarray}
\label{E-3-2}
\op{F}^{\pm}
=
\sum_{i=1}^N 
\big(
&&
D_{0}(i)\op{s}^{\pm}(i) 
+
D_{\mp1}(i)\op{s}^{z}(i)
\\
&&
+ 
D_{\mp2}(i)\op{s}^{\mp}(i)
\big)
\nonumber
\ ,
\end{eqnarray}
where $D_{0}(i) =\alpha_i(3\cos\theta_i -1)$,
$D_{\pm1}(i)=\alpha_i \sin\theta_i \cos\theta_i\\ \exp(\mp
i\varphi_i)$, $D_{\mp2}=1/2\alpha_i \sin^2\alpha_i
\exp(\mp2i\varphi_i)$ are the usual geometrical factors of the
dipolar interaction, $\alpha_i=3\gamma_N \gamma_S/(2r_i^3)$.
$\theta_i$ and $\varphi_i$ are the polar coordinates of the
vector $\vec{r}$ describing the relative positions of the two
spins. In the following we assume an isotropic case with
$\varphi$=0 and $\alpha_i$=1. $\gamma_S$ and $\gamma_N$ are the
gyromagnetic ratios of the electronic and nuclear spins,
respectively.

The spin-lattice relaxation is a resonant process
\cite{Abr:CO61} which ideally should only occur if the
transition energy $E_{\mu}-E_\nu$ in the spin system matches the
nuclear Lamor frequency. Nevertheless, the interaction of the
whole system with its surrounding broadens levels. In addition
the experimental resolution is limited. We therefore allow
transitions which deviate up to $\varepsilon$ from strict energy
conservation. This is taken care of by a Gaussian distribution
function $\delta_\varepsilon(\omega_N - \frac{E_{\mu}-E_\nu
}{\hbar})$.  This function could in principle depend both on
temperature and on applied field \cite{BLL:PRB04,SCL:PRL05}. We
will neglect such possible dependencies and use the same
function with $\varepsilon=0.2$~K for all calculations.  The
interested reader is referred to
Refs.~\cite{BLL:PRB04,SCL:PRL05}, where possible temperature and
field dependencies are discussed.

\begin{figure}[!ht]
\begin{center}
\includegraphics[clip,width=70mm]{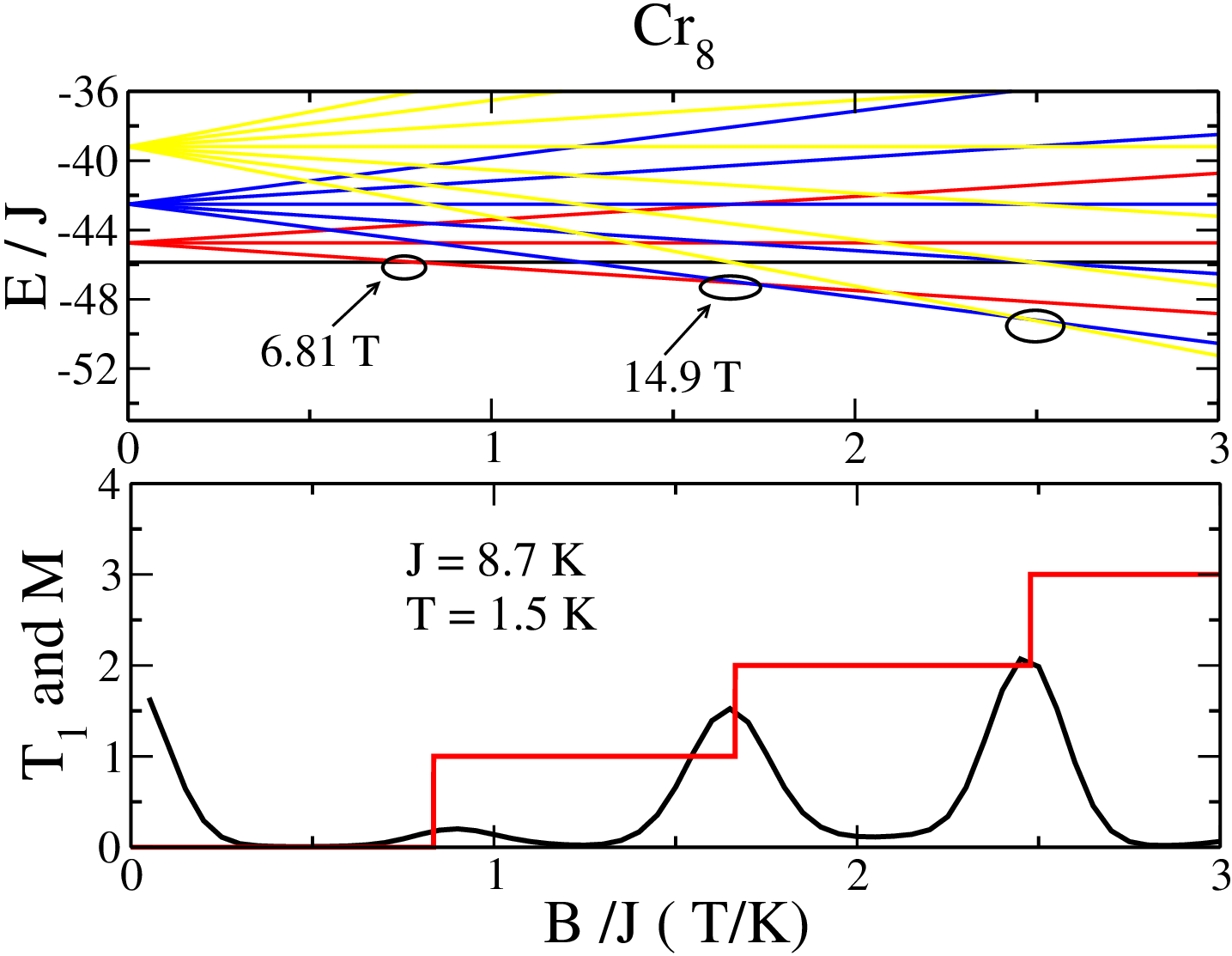}
\vspace*{1mm}
\caption[]{Top panel: Zeeman splitting of the low-lying levels
  of Cr$_8$. The crossing fields are highlighted and the values
  of the two lowest fields given.
  Bottom panel: $(T=0)$-magnetization (steps) and relaxation
  rate $T_1^{-1}$ as function of the applied field normalized to
  the coupling $J$.}
\label{F-L}
\end{center}
\end{figure}

The spin-lattice relaxation of the mother substance Cr$_8$ has
been investigated in great detail \cite{BLL:PRB04,SCL:PRL05},
but predominantly as a function of temperature for certain small
applied magnetic fields. In the following we discuss the
behavior of the relaxation rate as a function of magnetic field
for a typical small temperature of $T=1.5$~K \cite{LBJ:JMMM04}.
This function highlights the behavior of the magnetic system at
low-lying (dominantly ground state) Zeeman level crossings,
since there resonant cross relaxation occurs. Experimentally
such data are rarely accessible due to the fact that often the
level crossing fields are outside the producible field range. In
the case of Cr$_8$ \cite{LBJ:JMMM04} and Fe$_{10}$
\cite{JJL:PRL99} these data could nevertheless be measured
thanks to moderate exchange constants. One important result of
these measurements is that the values of the level crossing
fields for even-membered Heisenberg rings follow the Land\'e
interval rule \cite{JJL:PRL99}, which is nowadays understood as
rotational modes \cite{ScL:PRB01}, rotation of the N\'eel vector
\cite{And:PR52,WGC:PRL03} or tower of states
\cite{And:PR52,ADM:PRL93}.

\begin{figure}[!ht]
\begin{center}
\includegraphics[clip,width=70mm]{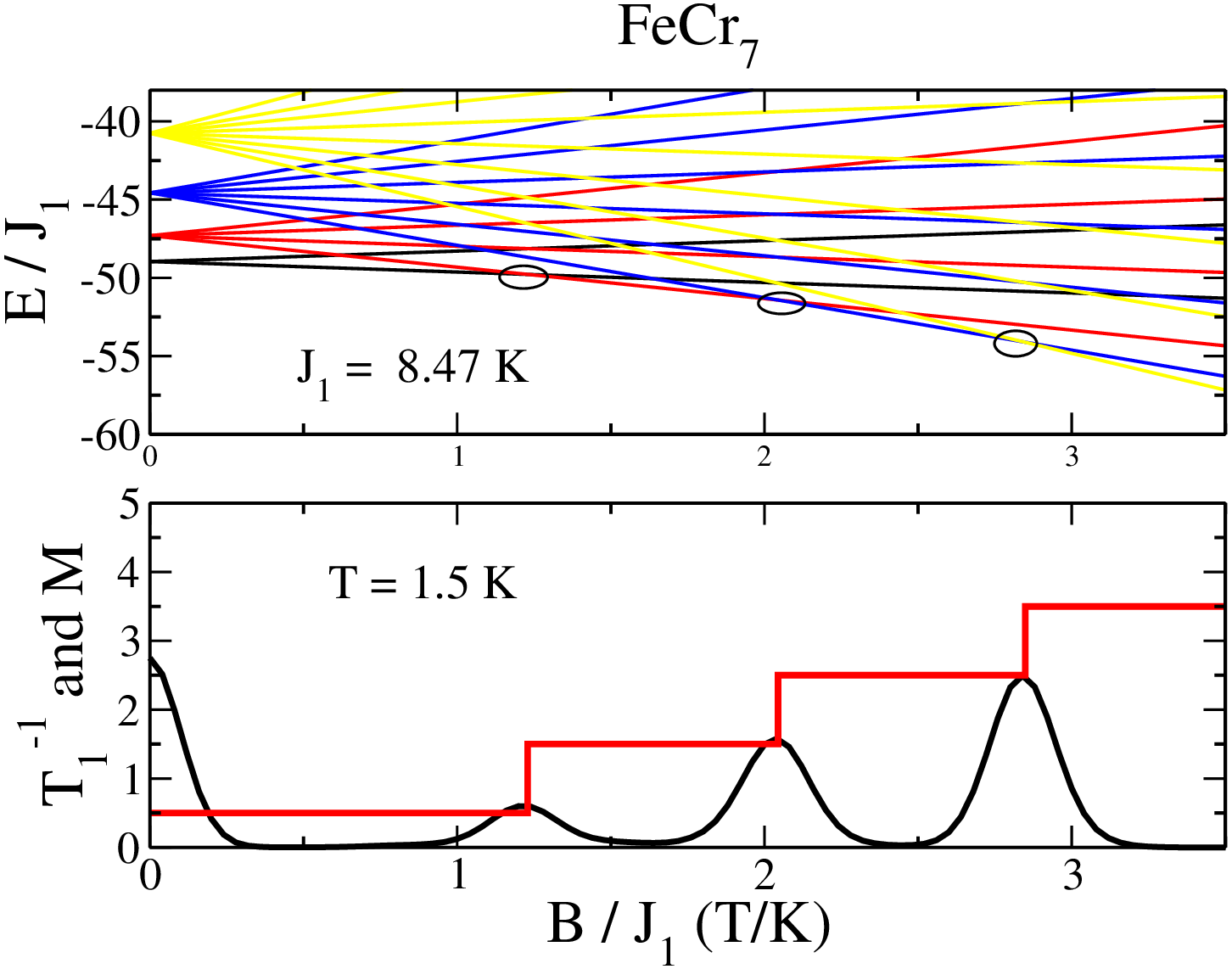}
\vspace*{1mm}
\caption[]{Top panel: Zeeman splitting of the low-lying levels
  of {FeCr$_7$}. The crossing fields are highlighted and the values
  of the two lowest fields given.
  Bottom panel: $(T=0)$-magnetization (steps) and relaxation
  rate $T_1^{-1}$ as function of the applied field normalized to
  the coupling $J_1$.}
\label{F-H}
\end{center}
\end{figure}

\begin{figure}[!ht]
\begin{center}
\includegraphics[clip,width=70mm]{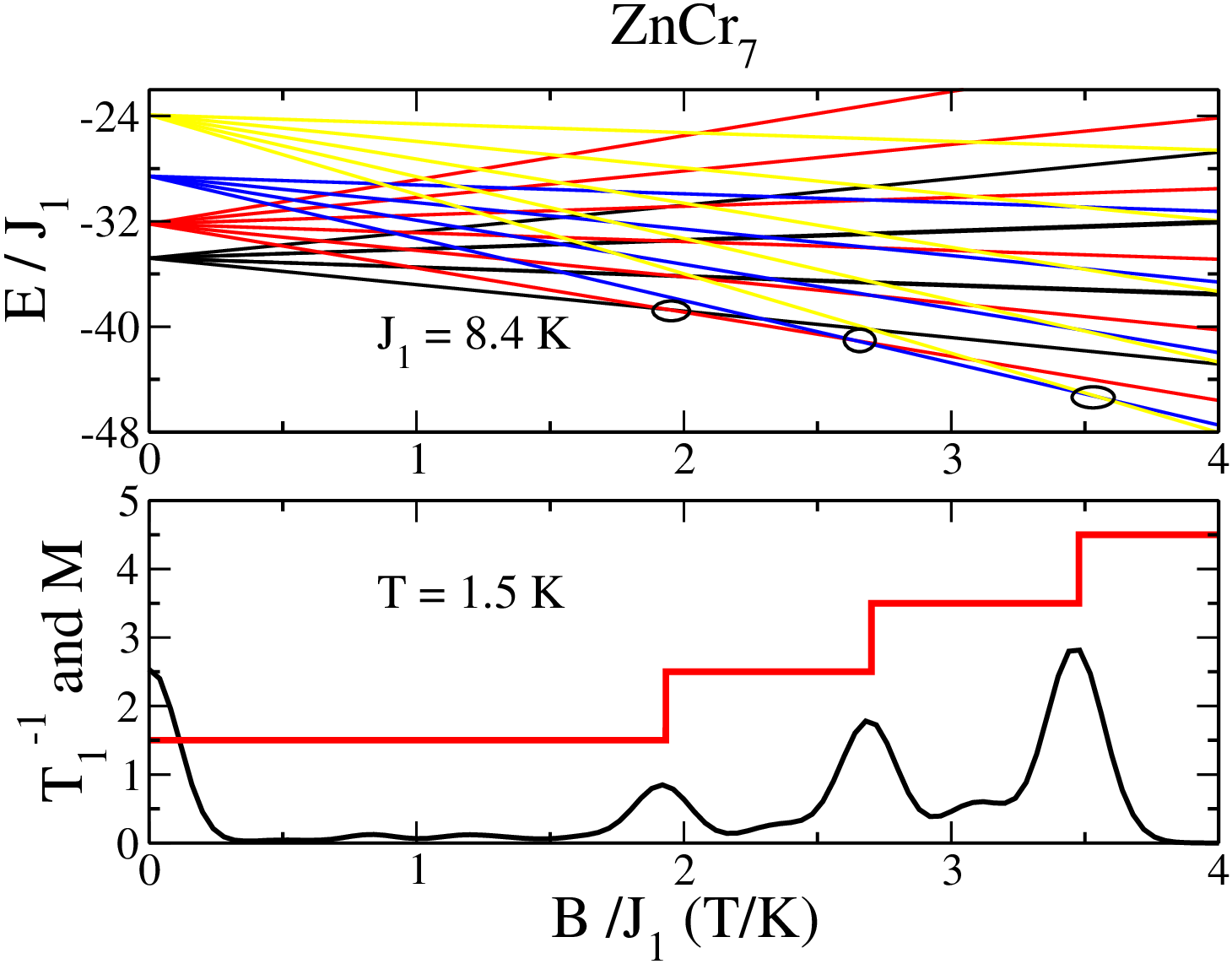}
\vspace*{1mm}
\caption[]{Top panel: Zeeman splitting of the low-lying levels
  of {ZnCr$_7$}. The crossing fields are highlighted and the values
  of the two lowest fields given.
  Bottom panel: $(T=0)$-magnetization (steps) and relaxation
  rate $T_1^{-1}$ as function of the applied field normalized to
  the coupling $J_1$.}
\label{F-K}
\end{center}
\end{figure}

\begin{figure}[!ht]
\begin{center}
\includegraphics[clip,width=70mm]{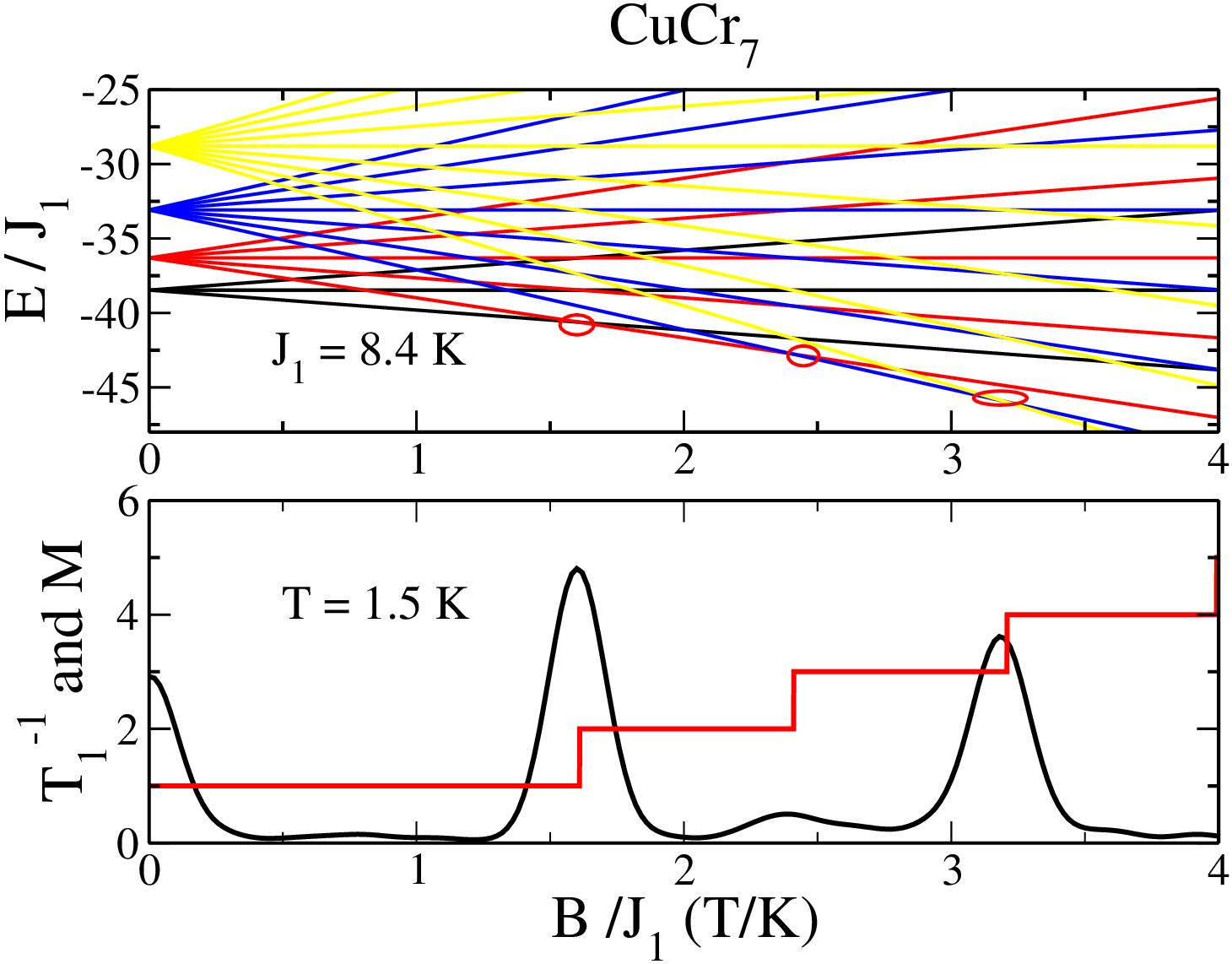}
\vspace*{1mm}
\caption[]{Top panel: Zeeman splitting of the low-lying levels
  of {CuCr$_7$}. The crossing fields are highlighted and the values
  of the two lowest fields given.
  Bottom panel: $(T=0)$-magnetization (steps) and relaxation
  rate $T_1^{-1}$ as function of the applied field normalized to
  the coupling $J_1$.}
\label{F-I}
\end{center}
\end{figure}

\begin{figure}[!ht]
\begin{center}
\includegraphics[clip,width=70mm]{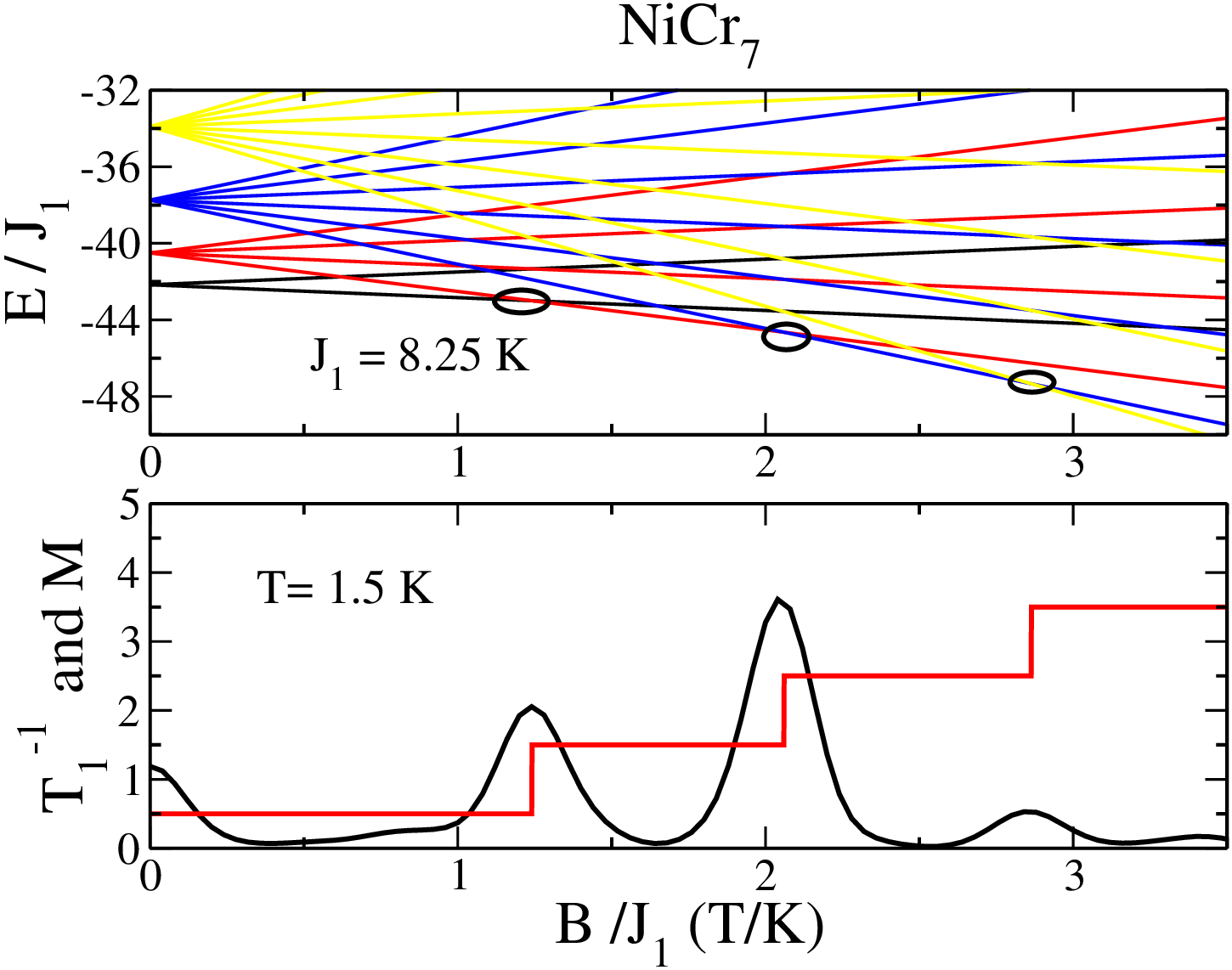}
\vspace*{1mm}
\caption[]{Top panel: Zeeman splitting of the low-lying levels
  of {NiCr$_7$}. The crossing fields are highlighted and the values
  of the two lowest fields given.
  Bottom panel: $(T=0)$-magnetization (steps) and relaxation
  rate $T_1^{-1}$ as function of the applied field normalized to
  the coupling $J_1$.}
\label{F-J}
\end{center}
\end{figure}

An obvious difference between Cr$_8$ and the heterometallic
{MCr$_7$} wheels is given by the fact that all of the discussed
wheels have ground states with non-vanishing total spin.
Therefore, for {FeCr$_7$} (\figref{F-H}), {ZnCr$_7$}
(\figref{F-K}), {CuCr$_7$} (\figref{F-I}), and {NiCr$_7$}
(\figref{F-J}) resonant relaxation occurs already at very low
magnetic fields, which expresses itself in the pronounced
maximum seen around $B=0$ in Figs.~\ref{F-H}-\ref{F-J}.

The second deviation from the behavior of Cr$_8$ consists in
pronounced differences of the maximum rates at higher Zeeman
level crossings in the cases of {CuCr$_7$} and {NiCr$_7$}.
Within the employed framework and the assumed approximations the
relaxation at the crossing between $S=1$ and $S=2$ in {CuCr$_7$}
(\figref{F-I}) appears to be rather small.  The same is true for
the relaxation at the crossing between $S=5/2$ and $S=7/2$ in
{NiCr$_7$} (\figref{F-J}).

\section{Summary and outlook}
\label{sec-5}

In this article we have reexamined the available experimental
susceptibility data \cite{LME:ACIE03} on heterometallic
{MCr$_7$} ring molecules in terms of a simple isotropic
Heisenberg Hamiltonian for M=Fe, Ni, Cu, and Zn.  Our main
results are that in the case of {FeCr$_7$} the iron-chromium
exchange is different from the chromium-chromium exchange in
contrast to the other cases and that for {CuCr$_7$} and
{NiCr$_7$} unexpectedly reduced proton spin-lattice relaxation
rates $T_1^{-1}$ occur at certain level crossings. It would be
very interesting to see whether this behavior could be
experimentally verified or whether the additional anisotropic
terms in the Hamiltonian
\cite{VSG:CEJ02,CVG:PRB03,AGC:PRB03,LME:ACIE03,CGA:04} alter the
picture completely.

\section*{Acknowledgement}

This work was supported by the Ph.D. program of the University
of Osnabr\"uck.  We would like to thank Eva Rentschler (Mainz)
for providing the susceptibility data and Bernd Pilawa
(Karlsruhe) as well as Richard Winpenny (Manchester) for
valuable discussions. We would also like to thank Alessandro
Lascialfari for informing us about upcoming NMR measurements at
{NiCr$_7$} accompanied by detailed theoretical calculations
including anisotropic terms.


\end{document}